\documentclass[11pt]{article}
\renewcommand{\theequation}{\arabic{section}.\arabic{equation}}
\def\be{\begin{equation}}
\def\ee{\end{equation}}
\def\ba{\begin{eqnarray}}
\def\ea{\end{eqnarray}}
\def\nn{\nonumber}
\def\lb{\label}
\newcommand{\bm}[1]{\mbox{\boldmath $#1$}}
\def\R{{\cal R}}
\def\bR{\overline{\cal R}}
\def\bT{\overline{T}}
\def\bL{\overline{L}}
\def\bE{\overline{E}}
\def\X{{\bf X}}
\def\L{{\bf L}}
\def\a{{\bm\alpha}} \def\b{{\bm\beta}} \def\c{{\bm\gamma}}
\def\1{{\bf 1}}
\def\bb{\bibitem}
\begin{document}

\begin{titlepage}
\date{}
\title{
\begin{flushright}\begin{small} LAPTH-026/25 \end{small} \end{flushright} \vspace{1cm}
Black holes and wormholes in sourceless\\ three-dimensional conformal Killing gravity}

\author{G\'erard Cl\'ement$^a$\thanks{Email: gclement@lapth.cnrs.fr} and
Khireddine Nouicer$^{b}$\thanks{Email: khnouicer@univ-jijel.dz} \\ \\
$^a$ {\small LAPTh, Universit\'e Savoie Mont Blanc, CNRS, F-74940  Annecy, France} \\
$^b$ {\small LPTh, Department of Physics, University of Jijel,} \\
{\small BP 98, Ouled Aissa, Jijel 18000, Algeria}}

\maketitle

\begin{abstract}
We derive all the sourceless solutions of three-dimensional conformal Killing gravity with two Killing vectors. Along with singular solutions and BTZ black holes, the stationary solutions include regular warped AdS3 black holes and wormholes.
\end{abstract}

\end{titlepage}
\setcounter{page}{2}

\section{Introduction}

Lower-dimensional gravity provides a simplified yet powerful framework for probing the foundational structure of gravitational theories. In particular, 
(2+1)-dimensional general relativity exhibits distinctive characteristics: the Riemann tensor is entirely determined by the Ricci tensor, and the 
theory possesses no local propagating degrees of freedom \cite{Carlip}. Despite this apparent simplicity, it admits rich and non-trivial solutions. 
A paradigmatic example is the rotating black hole solution discovered in 1992 by Ba${\tilde n}$ados, Teitelboim, and Zanelli ???- now known as the BTZ black hole 
\cite{btz}. This vacuum solution exists in a spacetime with a negative cosmological constant and mirrors many features of higher-dimensional black holes, 
including event and Cauchy horizons, thermodynamic behavior, and relevance in holographic dualities. Simultaneously, it retains unique traits such as 
locally AdS3 geometry and constant curvature invariants.

A variety of alternative theories of (2+1)-dimensional gravity, such as cosmological topologically massive gravity \cite{TMG} and new massive gravity 
\cite{NMG}, have been proposed. These theories generically admit two very different kinds of black hole solutions, the BTZ black holes, and
warped AdS3 black holes \cite{tmgbh,ALPSS,tmgebh}. Similarly to BTZ black holes, these have four Killing vectors generating the algebra 
$sl(2,R)\times R$, and constant curvature invariants. The ADM lapse function of these warped black holes goes to a constant value at spacelike infinity 
\cite{tmgbh}, which makes them closer in this respect to four-dimensional black holes. On the other hand, warped AdS3 black holes differ from both 
BTZ black holes and four-dimensional black holes in being intrinsically non-static, their ergosphere extending to infinity \cite{tmgbh}.

0ne of the most recent developments in alternative theories of four-dimensional gravity is oonformal Killing Gravity (CKG), a novel extension of 
general relativity proposed by J. Harada \cite{harada2023a}. In this theory, Einstein's field equations are replaced with differential equations 
of third order, with only one coupling constant, Newton's gravitational constant $G$. It was shown in \cite{mantica2023} that the Harada field 
equations are equivalent to the Einstein equations modified by the addition to the matter energy-momentum tensor of a divergence-free 
conformal Killing tensor. While by construction, all solutions of the Einstein equations are solutions of this new theory, CKG presents a broader 
spectrum of gravitational dynamics and accommodates new, genuinely non-Einsteinian solutions, both stationary and cosmological 
\cite{harada2023a,harada2023b,tarciso2023,mantica2024a,barnes2024a,HK,hervik2024,tarciso2024,gurses2024}. 

Our goal in this work is to assess the compatibility of CKG with lower-dimensional black hole geometries and 
to explore the possibility of novel three-dimensional solutions that go beyond those allowed in standard general relativity.
The paper is structured as follows: section 2 presents briefly the formulation of conformal Killing Gravity in three dimensions. 
In Section 3, we determine the general stationary circularly symmetric solution of this theory, which depends on nine integration constants. 
We then show in Section 4 that solutions free from naked singularities are either BTZ black holes and wormholes, or warped AdS3 black holes 
and wormholes, which we review in section 5. Our findings are summarized in the last section.

\setcounter{equation}{0}
\section{Harada's equations in $D$ dimensions}

We first extend the gravitational field equations originally proposed by Harada \cite{harada2023a} in four dimensions to the case of $D$-dimensional 
spacetime. Harada first wrote down the equations
\be\lb{ckg}
H_{\mu\nu\rho} =  8\pi GT_{\mu\nu\rho},
\ee
where the totally symmetric tensor $H_{\mu\nu\rho}$ is defined by
\begin{equation}\lb{H}
H_{\mu\nu\rho} \equiv a\left[D_{\mu}\R_{\nu\rho}+D_{\nu}\R_{\rho\mu}+D_{\rho}\R_{\mu\nu}\right]
+ b\left[g_{\nu\rho}\partial_{\mu}+g_{\rho\mu}\partial_{\nu}+g_{\mu\nu}\partial_{\rho}\right]\bR
\end{equation}
(with $\R_{\mu\nu}$ the Ricci tensor, $\bR$ its trace, and $D$ the covariant derivative), and
the matter three-tensor $T_{\mu\nu\rho}$ is similarly defined as:
\begin{equation}\label{T}
T_{\mu\nu\rho} \equiv c\left[D_{\mu}T_{\nu\rho}+D_{\nu}T_{\rho\mu}+D_{\rho}T_{\mu\nu}\right]
+ d\left[D_{\nu\rho}\partial_{\mu}+g_{\rho\mu}\partial_{\nu}+g_{\mu\nu}\partial_{\rho}\right]\bT,
\end{equation}
(with $T_{\mu\nu}$ the energy-momentum tensor, and $\bT$ its trace). He then determined the coefficients $a$, $b$, $c$, $d$
by requiring 1) that the conservation of energy-momentum $D_\nu T_\mu^\nu = 0$ is a consequence of the field equations (\ref{ckg}), and
2) that every solution of the Einstein equations $\R_{\mu\nu} - (1/2) \bR g_{\mu\nu} = T_{\mu\nu}$ is a solution of equations (\ref{ckg}). Contracting (\ref{ckg}) with $g_{\nu\rho}$ and using the contracted Bianchi identity $D_{\mu}\R^\mu_\nu - (1/2)\partial_{\nu}\bR \equiv 0$ leads to
\be
[2a + (D+2)b]\partial_{\mu}\bR = 2cD_\nu T_\mu^\nu + [c + (D+2)d]\partial_{\mu}\bT,
\ee
so that requirement 1) is satisfied provided
\be\lb{cond1}
2a + (D+2)b = 0, \quad c + (D+2)d = 0.
\ee
Similarly, requirement 2) leads to the conditions
\be\lb{cond2}
a - c = 0, \quad 2b + c + (D-2)d = 0.
\ee
Choosing the normalisation $a = 1$, the redundant system (\ref{cond1})-(\ref{cond2}) is solved by
\be
a = 1,\; b = - \frac2{D+2},\; c = 1,\; d = - \frac1{D+2}.
\ee

In two dimensions ($D=2$), the Einstein tensor $\R_{\mu\nu} - (1/2) \bR g_{\mu\nu}$ vanishes identically, and so does the Harada tensor $H_{\mu\nu\rho}$. In three dimensions ($D=3$), the Harada equations are (\ref{ckg}) with the curvature three-tensor
\begin{equation}\lb{H3}
H_{\mu\nu\rho} \equiv \left[D_{\mu}\R_{\nu\rho}+D_{\nu}\R_{\rho\mu}+D_{\rho}\R_{\mu\nu}\right]
- \frac25\left[g_{\nu\rho}\partial_{\mu}+g_{\rho\mu}\partial_{\nu}+g_{\mu\nu}\partial_{\rho}\right]\bR
\end{equation}
and the matter three-tensor
\begin{equation}\label{T3}
T_{\mu\nu\rho} \equiv \left[D_{\mu}T_{\nu\rho}+D_{\nu}T_{\rho\mu}+D_{\rho}T_{\mu\nu}\right]
- \frac15\left[g_{\nu\rho}\partial_{\mu}+g_{\rho\mu}\partial_{\nu}+g_{\mu\nu}\partial_{\rho}\right]\bT.
\end{equation}
The totally symmetric Harada tensor $H_{\mu\nu\rho}$ has ten components, constrained by the three Bianchi identities
\be\lb{bianchi}
g^{\mu\nu}H_{\mu\nu\rho} = 0,
\ee
leaving only seven algebraically independent Harada equations (\ref{ckg}).

\setcounter{equation}{0}
\section{Vacuum solutions with two Killing vectors}

We assume the existence of two commuting Killing vectors. and choose a coordinate system so that the metric can be written as:
\begin{equation}
ds^{2}=\lambda_{ab}(\rho)\,dx^{a}dx^{b}+\frac{1}{\zeta^2(\rho)R^{2}(\rho)}\,d\rho^{2}\,,\label{metric}
\end{equation}
($a,b = 0,1$), where $R^2 = \vert{\rm det}\lambda\vert$, and $\zeta(\rho)$
is a scale function. The metric (\ref{metric}) is invariant under reparameterizations of the radial coordinate $\rho$,
which allows to fix $\underline{\zeta=1}$. The only nonzero components of the Harada tensor are $H_{222}$ and the three $H_{ab2}$
(with $x^2 = \rho$), constrained by the Bianchi identity
\be\lb{bianchi1}
g^{ab}H_{ab2} + g^{22}H_{222} \equiv 0,
\ee
so that there are three independent third-order differential Harada equations (\ref{ckg}) for the three components
of the symmetric matrix $\lambda_{ab}(\rho)$. In the sourceless case, we can take these to be the scalar equation
\be\lb{scalar}
H_{222}=0
\ee
and the traceless matrix equation
\be\lb{traceless}
H_{ab2} - (1/2)g_{ab}g^{cd}H_{cd2} = 0.
\ee

The metric (\ref{metric}) is also invariant  under SL($2,R$) transformations in the 2-space spanned by the $x^a$. The local isomorphism
${\rm SL}(2,R) \approx {\rm SO}(2,1)$ suggests the parametrization \cite{EL}
\begin{equation}
\lambda=\left(
\begin{array}
[c]{cc} T+X & Y\\ Y & T-X
\end{array}
\right)  ,\label{lambda}
\end{equation}
where the vector ${\bf X} = (T,\,X,\,Y)$ belongs to a superspace endowed with a Minkowskian metric
$\eta_{ij} = {\rm diag}(-1,1,1)$ ($i,j=0,1,2$). On this superspace, we define the dot, wedge and mixed products of the
supervectors $\bf A$, $\bf B$ and $\bf C$ by
\be\lb{dotwedge}
{\bf A}\cdot{\bf B} = \eta_{ij}A^{i}B^{j}, \quad ({\bf A}\wedge{\bf B})^{i} = \eta^{ij}\epsilon_{jkl}A^{k}B^{l}, \quad ({\bf A},{\bf B},{\bf C}) = ({\bf A}\wedge{\bf B})\cdot{\bf C}\,,
\ee
where $\epsilon_{ijk}$ is the Levi-Cevita symbol, with $\epsilon_{012}=+1$.
The inverse metric function $R^2$ in (\ref{metric}) is then the norm
\be
R^2 \equiv {\bf X}^2 = -T^2 +X^2 +Y^2\,.
\ee

Using this parametrization, we obtain the mixed Ricci tensor components (see Appendix A)
\be
{\R^a}_b = - \frac12\bigg((\X\cdot\X')'\1 +\ell'\bigg)_{\ \ b}^{a}\,,
\quad \R_{\ \ 2}^{2}= -(\X\cdot\X')' + \frac{1}{2}{\X'}^2\,,
\ee
where the prime stands for the derivative $d/d\rho$, $\1$ is the unit matrix, and $\ell$ is the traceless matrix
\be\lb{ell}
\ell = [\L] = \left(
\begin{array}
[c]{cc} -L^{Y} & L^X - L^T \\ L^X + L^T & L^{Y}
\end{array}\right),
\ee
associated with the super-angular momentum vector
\be
\L \equiv \X\wedge\X'\,.
\ee

The computation of the component ${H^2}_{22}$ is straightforward:
\ba\lb{scalar1}
\frac13\,{H^2}_{22} &=& D_2 \R^2_2 - \frac25\partial_2\bR = \left(\R^2_2 - \frac25\bR\right)' \nn\\
&=& \frac15\left(3\R^2_2 - 2\R^a_a\right)' = - \frac15\,\X\cdot\X'''\,.
\ea
The scalar equation (\ref{scalar}) is thus simply
\be\lb{scalar2}
\X\cdot\X''' = 0.
\ee

The matrix components ${H^a}_{b2}$ are evaluated in Appendix A. The traceless matrix equation (\ref{traceless}) leads to\footnote{The scalar equation (\ref{scalar1}) has not been used.}
\be\lb{traceless1}
- (\X^2)\,{\ell}'' + (\X\cdot\X')\,{\ell}' -  (\X\cdot\X'')\,{\ell} - [\L\wedge\L']  = 0,
\ee
which translates into the vector equation
\be\lb{vector}
(\X^2)\,\L'' - (\X\cdot\X')\,\L' + (\X\cdot\X'')\,\L + \L\wedge\L' = 0.
\ee
Using the definition $\L = \X\wedge\X'$ and the derivatives $\L' = \X\wedge\X''$, $\L'' = \X\wedge\X''' + \X'\wedge\X''$, (\ref{vector}) can be rewritten
\be\lb{vector1}
(\X^2)[\X\wedge\X''' + \X'\wedge\X''] - (\X\cdot\X')\X\wedge\X'' + (\X\cdot\X'')\X\wedge\X' = (\X,\X',\X'')\X.
\ee

The vector product of (\ref{vector1}) with $\X$ leads to the simple consequence
\be\lb{X3}
(\X^2)\X''' - (\X\cdot\X''')\X = 0.
\ee
On account of (\ref{scalar2}) and of $(\X^2)= R^2 \neq0$, this means that
\be\lb{X31}
\X''' = 0,
\ee
which is integrated by
\be\lb{quadra}
\X = \a\rho^2 + \b\rho + \c,
\ee
where $\a$, $\b$, $\c$ are three constant vectors. The reduced vector equation
\be\lb{vector2}
(\X^2)\X'\wedge\X'' + (\X\cdot\X')\X''\wedge\X + (\X\cdot\X'')\X\wedge\X' = (\X,\X',\X'')\X
\ee
is identically satisfied, so that (\ref{quadra}) is the general stationary circularly symmetric solution of three-dimensional conformal Killing gravity.

Returning to the matrix parametrization, we have found that the four apparently independent equations (\ref{scalar}) and (\ref{traceless}) reduce to the three third-order differential equations for the symmetric $2\times2$ matrix $\lambda(\rho)$
\be
\lambda'''(\rho) = 0,
\ee
which are solved by
\be\lb{quadra1}
\lambda = A\rho^2 + B\rho + C,
\ee
where $A$, $B$, $C$ are three arbitrary constant symmetric matrices (nine integration constants).

\setcounter{equation}{0}
\section{Geometry}

Let us now discuss the spacetime geometry of the solution (\ref{quadra}), depending on the characteristics of the constant matrices $A$, $B$, $C$, or equivalently of the constant three-vectors $\a$, $\b$, $\c$. In the following, we assume that the Killing vector $\partial_0 = \partial_t$ is time-like for $\rho\to+\infty$, while the other Killing vector $\partial_1 = \partial_{\varphi}$ has closed orbits and is space-like for $\rho\to+\infty$.

For $\a=0$, (\ref{quadra}) reduces to the general stationary rotationally symmetric solution of three-dimensional general relativity with cosmological constant \cite{EL}, including the well-known BTZ solution \cite{btz}
\ba
ds^2 &=& (M/2 - 2\ell^{-2}\rho)dt^2 - Jdtd\varphi + (2\rho + M\ell^{2}/2)d\varphi^2 \nn\\
&& + [4\ell^{-2}\rho^2 + (J^2 - M^2\ell^2)/4]^{-1}d\rho^2,
\ea
describing black holes for $J^2 \le M^2\ell^2$, and Lorentzian wormholes \cite{EL} in the over-rotating case $J^2 > M^2\ell^2$.

So we discuss only the non-trivial case $\underline{\a\neq0}$. The Ricci scalar
\be
\R = -  2 \X\cdot\X'' - \frac32\X'^2 = - 10\a^2\rho^2 - 10\a\cdot\b\rho - 4\a\cdot\c - \frac32\b^2                                                                                      \ee
is everywhere finite, but generically diverges at infinity. So there is a naked curvature singularity at $\rho\to+\infty$, unless $\a^2 = \a \cdot\b = 0$.

The first-integrated geodesic equations for the metric (\ref{metric}) are
\ba\lb{geo}
&\dot{x}^a = \lambda^{ab}\Pi_b, & \nn\\
&\dot{\rho}^2 + W(\rho) = 0, &
\ea
where the dot denotes the derivative $d/d\tau$ with respect to an affine parameter $\tau$ along the geodesic, $\lambda^{ab}$ is the matrix inverse of $\lambda_{ab}$, $\Pi_a$ are constants of the motion, and the effective potential  is
\be\lb{pot}
W(\rho) = R^2 \Pi_a\lambda^{ab}\Pi_b - \epsilon R^2,
\ee
where $\epsilon = -1, 0, +1$ for timelike, null, or spacelike geodesics. The matrix $R^2\lambda^{-1}$ is, as the matrix $\lambda$, generically quadratic in $\rho$, while
\be\lb{R2}
R^2 = \a^2\rho^4 + 2\a\cdot\b\rho^3 + (2\a\cdot\c + \b^2)\rho^2 + 2\b\cdot\c\rho + \c^2
\ee
is quartic.

The possible horizon structure depends on the degree of $R^2(\rho)$. If $R^2$ is quadratic in $\rho$ (generic case), there can be up to four horizons. However, outside the outermost horizon, there is a naked singularity at $\rho\to+\infty$. So, although these spacetimes present horizons, they are not black-hole spacetimes. If $R^2(\rho)$ is positive definite, there is no horizon, and the spacetime ends on two naked singularities at $\rho\to\pm\infty$.

If $\a^2 = 0$, $R^2$ is cubic in $\rho$. There are then between one and three horizons. Again, the spacetime ends on a naked singularity at $\rho\to+\infty$.

So the only non-trivial, non-singular case is for
\be\lb{wads3}
\a^2 = \a \cdot\b = 0.
\ee
In this case, the three-dimensional geometry is of the warped AdS3 type \cite{ALPSS}. This geometry has been found to solve the equations of 
topologically massive gravity \cite{tmgbh,adtmg}, of topologically massive gravito-electrodynamics \cite{tmgebh}, of new massive gravity 
\cite{newmass}, and some other three-dimensional theories of gravity \cite{nam2010,tonni2010,ghodsi2011a,ghodsi2011b}. In the present case, 
the geometry is unconstrained by model parameters, which prompts us to reexamine in the next section, generalizing the analysis of \cite{tmgebh}, 
the most general possible structures of the solution (\ref{quadra}) under the constraints (\ref{wads3}).

\setcounter{equation}{0}
\section{Warped AdS3 black holes and wormholes}

As discussed in \cite{tmgebh}, the constraints  (\ref{wads3}) on the Lorentzian three-vectors $\a$ and $\b$ imply
\be\lb{d}
\a\wedge\b = c\a\,, \qquad \b^2 = c^2\,,
\ee
for some real constant $c$. In the following, we choose a length scale so that $c=1$. Also, we can choose a rotating frame such that the mixed component $\lambda_{01}$ of the metric is only linear in $\rho$. The vectors $\a$, $\b$ and $\c$ can then be parameterized by
\be\lb{para}
\a = (a,\pm a,0)\,,\quad \b = (b,\pm b,\mp 1)\,, \quad \c = (z+u,z-u,v)\,,
\ee
The metric written in ADM form
\be\lb{adm}
ds^2 = - \frac{R^2}{r^2}dt^2 + r^2(d\varphi - \omega dt)^2 + \frac{d\rho^2}{R^2},
\ee
with
\be\lb{omega}
r^2 = T-X, \quad \omega = - \frac{Y}{T-X} = \pm\frac{\rho\mp v}{r ^2},
\ee
is stationary if $R^2(\infty)>0$, and causally regular if $r^2$ is positive when $R^2$ is positive.

We consider in the following the two possible signs in (\ref{para}) (only the lower sign was considered in \cite{tmgebh}).

\subsection{Lower sign}
In this case,
\be
r^2 = 2(a\rho^2 + b\rho + u),
\ee
while
\be
R^2 = (1-4az)\rho^2 - 2(v+2bz)\rho + v^2 - 4uz.
\ee

If the quadratic term vanishes, $1-4az=0$, the metric represents a black hole with a single horizon. If both the quadratic and linear term vanish, the metric is horizonless. We refer the reader to \cite{tmgebh} for further discussions of these cases, and consider now the generic case where $R^2$ is quadratic in $\rho$ and the quadratic term is positive for stationary solutions. $R^2$ can either have two roots, corresponding to a black hole with two horizons, a double root, corresponding to an extreme black hole, or no root, in which case the geometry is of the wormhole type. We discuss first the black hole (or extreme black hole) case, following \cite{tmgebh}, then the wormhole case.

\subsubsection{Black holes}
The linear term may be set to zero ($v=-2bz$) by a translation on $\rho$, leading to
\be\lb{R2black}
R^2 = \beta^2(\rho^2-{\rho_0}^2),
\ee
with $\beta^2=1-4az$. Next, we observe that the lapse $N^2 = R^2/r^2$ goes for $\rho\to\infty$ to a constant value, which may be set to $1$ by rescaling the time coordinate $t$, so that $a = \beta^2/2$, $z = (1-\beta^2)/2\beta^2$.
The warped AdS3 black hole solution is thus (\ref{adm}) with $R^2$ given by (\ref{R2black}), $\omega$ given by (\ref{omega}), and\footnote{We assume here $\beta^2\neq1$. The limiting case $\beta^2=1$ could be treated as in \cite{tmgebh}.}
\be
r^2 = \beta^2\left[\rho^2 - \frac{2v}{1-\beta^2}\rho + \frac{v^2+\beta^2{\rho_0}^2}{1-\beta^2}\right].
\ee

The solution depends apparently on the three parameters $\beta$, $\rho_0$, $v$. However there is a fourth free parameter, which is the period of the cyclic variable $\varphi$. The geometry being free from conical singularities for all values of $\rho$, including at infinity, this period is completely arbitrary.

The possible zeros of $r^2(\rho)$ are located at
\be
\rho_{\pm} = \frac{1}{1-\beta^2}\left[v \pm \beta\sqrt{v^2 - (1-\beta^2){\rho_0}^2}\right] .
\ee
So closed timelike curves (CTC) occur between these two limits, unless $\beta^2 < 1$ and
$ v^2<(1-\beta^2){\rho_0}^2$. If $\beta^2 < 1$ and $ v^2>(1-\beta^2){\rho_0}^2$,  CTC do occur
for $\rho_-<\rho<\rho_+$, but are safely hidden behind the two horizons
($\rho_- < \rho_+ < -\rho_0$) if $v<0$.

\subsubsection{Wormholes}
This case is obtained from the black hole case by replacing $\rho_0$ by $i\mu$:
\ba
R^2 &=& \beta^2(\rho^2+\mu^2), \nn\\
r^2 &=& \beta^2\left[\rho^2 + \frac{2v}{\beta^2-1}\rho + \frac{\beta^2\mu^2-v^2}{\beta^2-1}\right],
\ea
where we have again chosen the time scale so that $N^2(\infty)=1$. The condition for the absence of CTC is now
$v^2 - (\beta^2-1)\mu^2 < 0$, so that the geometry is regular for all real $\rho$ if $\beta^2 > 1$ and  $v^2<(\beta^2-1)\mu^2$.

In this parameter range, the solution therefore represents a wormhole with two ends $\rho\to\pm\infty$. We now inquire whether this wormhole is traversable. Putting $\Pi_0=-E$, $\Pi_1=L$, the effective potential (\ref{pot}) for test particles ($\epsilon=-1$) is
\be
W(\rho) = \beta^2(1-E^2)\rho^2 + 2E\bL\rho + \frac{\beta^2-1}{\beta^2}\bL^2 - \alpha\beta^2\mu^2,
\ee
with $\bL = L - \beta^2Ev/(\beta^2-1)$, $\alpha = \beta^2E^2/(\beta^2-1)-1$. The wormhole is traversable for those parameter values for which the effective potential is negative definite, which is ensured for
\be
E^2 > 1, \quad \bL^2 < \frac{\beta^4\mu^2}{\beta^2-1}.
\ee

From (\ref{geo}), the angular proper velocity
\be
\dot\varphi = R^{-2}\left[E\rho + \frac{\beta^2-1}{\beta^2}\bL\right]
\ee
is everywhere finite and goes to zero as $\rho^{-1}$ at the two wormhole ends $\rho\to\pm\infty$. The radial proper velocity $\dot\rho$ being asymptotically proportional to $\rho$, it follows that the test particle makes a finite number of turns when going from one of the wormhole ends to the other.

\subsection{Upper sign}
This case can be obtained from the preceding by the interchange $t \leftrightarrow \varphi$. Now
\be
r^2 = 2u = {r_0}^2
\ee
is constant, and
\be\lb{R2up}
R^2 = (1-a{r_0}^2)\rho^2 - 2(v+b{r_0}^2)\rho + v^2 - 2z{r_0}^2  .
\ee

Again, the generic case where $R^2$ is quadratic can correspond either to a black hole or a wormhole. After carrying out a translation on $\rho$, the metric
\be
ds^2 = -\beta^2\frac{\rho^2-{\rho_0}^2}{{r_0}^2}dt^2 + {r_0}^2\bigg[d\varphi
  - \frac{\rho-v}{{r_0}^2}dt\bigg]^2 + \frac{d\rho^2}{\beta^2(\rho^2-{\rho_0}^2)}
\ee
represents a black cylinder with two horizons. This is regular for all values of the four real parameters $r_0$, $\rho_0$, $\beta$, $v$.

The analytic continuation $\rho_0 \to i\mu$ leads to the cylindrical wormhole
\be
ds^2 = -\beta^2\frac{\rho^2+\mu^2}{{r_0}^2}dt^2 + {r_0}^2\bigg[d\varphi
  - \frac{\rho-v}{{r_0}^2}dt\bigg]^2 + \frac{d\rho^2}{\beta^2(\rho^2+\mu^2)}.
\ee
The effective potential for test particles is
\be
W(\rho) = \left[\beta^2 - (1-\beta^2)\frac{L^2}{{r_0}^2}\right]\rho^2 + 2\bE L\rho - \bE^2{r_0}^2 + \beta^2\mu^2\left(1+\frac{L^2}{{r_0}^2}\right),
\ee
with $\bE = E - Lv/{r_0}^2$. This is negative at infinity for
\be
\beta^2 < 1, \quad L^2 > \frac{\beta^2}{1-\beta^2}{r_0}^2.
\ee
However the discriminant
\be
\Delta = \beta^2(L^2+{r_0}^2)\left[\bE^2 + \frac{\mu^2}{{r_0}^4}(1-\beta^2)\left(L^2-\frac{\beta^2}{1-\beta^2}{r_0}^2\right)\right]
\ee
is then positive. It follows that the effective potential always has two zeroes, between which it is positive, so that the wormhole is not traversable by test particles. The same conclusion holds for null geodesics ($\epsilon=0$), for which the effective potential is
\be
W(\rho) =  - (1-\beta^2)\frac{L^2}{{r_0}^2}\rho^2 + 2\bE L\rho - \bE^2{r_0}^2 + \beta^2\mu^2\frac{L^2}{{r_0}^2}.
\ee
This is negative at infinity provided $\beta^2 < 1$, in which case the effective potential again always has two zeroes.

If the quadratic term in (\ref{R2up}) vanishes, $a = 1/{r_0}^2$, the metric (where we have set $v=0$ by a translation on $\rho$)
\be
ds^2 = -\frac{\lambda(\rho-\rho_0)}{{r_0}^2}dt^2 + {r_0}^2\bigg(d\varphi
  - \frac{\rho}{{r_0}^2}dt\bigg)^2 + \frac{d\rho^2}{\lambda(\rho-\rho_0)}
\ee
represents a black cylinder with a single horizon. If both the quadratic and the linear term vanish, $R^2$ has a constant value which
we can set equal to ${r_0}^2$ by a time rescaling. After again setting $v=0$ by a translation on $\rho$, the metric
\be
ds^2 = - dt^2 + {r_0}^2\bigg(d\varphi
  - \frac{\rho}{{r_0}^2}dt\bigg)^2 + \frac{d\rho^2}{{r_0}^2}
\ee
represents a cylindrical Lorentzian wormhole. The effective potential for timelike geodesics is
\be
W(\rho) =  L^2 + {r_0}^2  - \frac1{{r_0}^2}\left(L\rho-E{r_0}^2\right)^2.
\ee
This has generically two zeroes, but is negative definite for $L=0$ and $E^2>1$. So this wormhole is traversable by radial timelike geodesics.

\section{Conclusion}

We have determined analytically the general sourceless solution of three-dimensional conformal gravity with two Killing vectors. In the 
parametrization of (\ref{metric}), the matrix elements $\lambda_{ab}(\rho)$ are quadratic in $\rho$. Imposing the absence of singularities 
at $\rho\to\pm\infty$, the resulting metric is either a solution of three-dimensional general relativity with cosmological constant, or of 
the warped AdS3 type. We have reviewed all the possible stationary rotationally symmetric WAdS3 metrics, which include, along with the 
WAdS3 black holes discussed in \cite{ALPSS} and \cite{tmgebh}, a class of black cylinders, possibly free from causal singularities, and 
two classes of Lorentzian wormholes. The wormholes of the first class are traversable, while those of the second class are generically not traversable.

\renewcommand{\theequation}{A.\arabic{equation}}
\setcounter{equation}{0}
\section*{Appendix}

We outline here the computation of the components of the Harada three-tensor using the superspace approach introduced in \cite{EL} and streamlined in \cite{adtmg}. To each supervector $\bf A$, we associate a traceless matrix noted
$[{\bf A}]$ or $a$ such that
\be\lb{vecmat}
a \equiv [\bf{A}]  = \left(
\begin{array}
[c]{cc} -A^{Y} & -A^{-}\\ A^{+} & A^{Y}
\end{array}\right),
\ee
where
\be
A^{\pm} \equiv A^T\pm A^X\,.
 \ee
For any two supervectors $\bf{A}$, $\bf{B}$, the product of the associated traceless matrices is
\be
[\bf{A}][\bf{B}] = (\bf{A}\cdot\bf{B})\,\1+[\bf{A}\wedge\bf{B}]\,, \label{product}
\ee
where $\1$ is the $2\times2$ unit matrix, and the Minkowskian dot and wedge products are defined in (\ref{dotwedge}).

From the metric (\ref{metric}) with $\zeta = 1$, we find the Christoffel symbols
\begin{equation}
\Gamma_{2b}^{a}=\frac{1}{2}\left(
\lambda^{-1}{\lambda}^{\prime}\right) _{\ \
b}^{a}\,,\quad\Gamma_{ab}^{2}=-\frac{1}{2}R ^{2}\emph{\
}{\lambda}^{\prime}{}_{ab}\,,\quad\Gamma_{22}^{2}=-R
^{-1}\,R^{\prime}\emph{\ }, \label{chris}
\end{equation}
where the prime stands for the derivative $d/d\rho$. The matrices in
(\ref{chris}) are related to the matrix $x$ (defined according to
(\ref{vecmat})) by
\be\lb{lamx}
\lambda =\tau^{0}x\,, \quad \lambda^{-1} = - \frac{1}{R^{2}}x\tau^{0}\,.
\ee
It follows that
\be
\lambda^{-1}{\lambda}^{\prime} =  R^{-2}xx' =  R^{-2}\left[(\X\cdot\X')\1 + \ell\right],
\ee
where $\ell$ is the matrix associated with the super-angular momentum vector
\be
\L \equiv \X\wedge\X^{\prime}\,.
\ee
This leads to the Ricci tensor components
\be
{\R^a}_b = -\frac1{2}\bigg(({RR}^{\prime})^{\prime}{\bf 1}+\ell^{\prime}\bigg)_{\ \ b}^{a}\,,
\quad \R_{\ \ 2}^{2}= -({RR}^{\prime})^{\prime}+\frac{1}{2}{\X}^{\prime2}\,,
\ee

The computation of the component ${H^2}_{22}$ of the Harada tensor has been carried out in the main text (equation (\ref{scalar1})). The computation of the matrix components is more involved. The definition (\ref{H3}) leads to
\ba
{H^a}_{b2} &=& D_2{\R^a}_{b} + \lambda^{ac}D_c\R_{b2} + D_b{\R^a}_{2} - \frac25\delta^a_b\partial_2\bR \nn\\
&=& \left(\R' - \R\lambda^{-1}\lambda' + \lambda^{-1}\lambda'\R_{\ \ 2}^{2} - \frac25\partial_2\bR\,\1\right)^a_{\ \ b}  \nn\\
&=& \frac12\bigg(- \ell''  + R^{-2}[\ell'\ell + (\X\cdot\X')\ell'- (\X\cdot\X')'\ell  + \X'^2\ell] + A\1
\bigg)^a_{\ \ b} \nn\\
&=& \frac12R^{-2}\bigg(- \X^2\ell'' + (\X\cdot\X')\ell' - (\X\cdot\X'')\ell - [\L\wedge\L'] + B\1 \bigg)^a_{\ \ b}\nn\\,
\ea
where we have used repeatedly (\ref{product}), and all the terms contributing to the trace ${H^c}_{c2}$ are grouped into the coefficients $A$, $B$.

\end{document}